# Diffusion-Inversion-Net (DIN): An End-to-End Direct Probabilistic Framework for Characterizing Hydraulic Conductivities and Quantifying Uncertainty

**Xun Zhang[1], Weijie Yang[2], Jiangjiang Zhang[3, 4], Simin Jiang[1]**

[1]College of Civil Engineering, Tongji University, Shanghai, China.

[2]School of Information, University of California, Berkeley, United States.

[3]Yangtze Institute for Conservation and Development, Hohai University, Nanjing, China.

[4]The National Key Laboratory of Water Disaster Prevention, Hohai University, Nanjing, China.

Corresponding author: Simin Jiang (jiangsimin@tongji.edu.cn)

**Key Points:**

- A deep learning framework is proposed for direct probabilistic inversion of hydraulic conductivity fields.
- The framework utilizes a diffusion model as a powerful geological prior learner, flexibly incorporating sparse, multi-source observations via conditional injection mechanisms.
- The framework directly generates posterior ensembles without iterative forward simulations, enabling effective uncertainty quantification.

**Abstract**

We propose the Diffusion-Inversion-Net (DIN) framework for inverse modeling of groundwater flow and solute transport processes. DIN utilizes an offline-trained Denoising Diffusion Probabilistic Model (DDPM) as a powerful prior leaner, which flexibly incorporates sparse, multi-source observational data, including hydraulic head, solute concentration, and hard conductivity data, through conditional injection mechanisms. These conditioning inputs subsequently guide the generative inversion process during sampling. Bypassing iterative forward simulations, DIN leverages stochastic sampling and probabilistic modeling mechanisms to directly generate ensembles of posterior parameter fields by repeatedly executing the reverse denoising process. Two representative posterior scenarios, Gaussian and non-Gaussian, are investigated. The results demonstrate that DIN can produce multiple constraint-satisfying realizations under identical observational conditions, accurately estimate hydraulic-conductivity fields, and achieve reliable uncertainty quantification. The framework exhibits strong generalization capability across diverse data distributions, offering a robust and unified alternative to conventional multi-stage inversion methodologies.

**Plain Language Summary**

Understanding the structure of subsurface aquifer is crucial for managing groundwater resources and predicting how pollution might spread, but it's difficult due to sparse borehole data. Traditional methods are extremely slow, requiring thousands of numerical model simulations. While newer and more complex inversion frameworks, which leverage deep learning models, are faster, they require the integration of multiple separate models. Both approaches struggle to accurately characterize complex geological structures, like fluvial channels. We developed a new AI method, Diffusion-Inversion-Net (DIN), that solves this problem directly. After a one-time offline training to learn geological patterns, DIN takes sparse field measurements (like water levels and pollution data) and generates a whole set of detailed, realistic maps that fit the data. Because our method provides this diverse ensemble, rather than just one guess, it also allows scientists to visualize uncertainty. This makes mapping the subsurface geological structures accurate.

**1 Introduction**

Accurate simulation of groundwater flow, solute transport, and characterization of aquifer heterogeneity are essential for effective water resources management and environmental protection (Kitanidis, 2015; Rizzo & de Barros, 2019). In recent years, deep learning has achieved remarkable success in hydrogeological modeling, enabling the data-driven characterization of highly complex and nonlinear systems that are often intractable using traditional numerical methods, particularly in solving inverse problems (Triplett et al., 2025; Wang et al., 2021). Groundwater inverse problems, defined as the inference of high-dimensional subsurface properties (e.g., hydraulic conductivity $K$ fields) from sparse and noisy observations (e.g., hydraulic head and contaminant concentrations), are typically characterized by high nonlinearity and ill-posedness. These challenges, which stem from strong subsurface heterogeneity and high-dimensional parameter spaces, are further compounded by substantial computational costs and difficulties in uncertainty quantification. Therefore, within the vision for developing hydrogeological Artificial General Intelligence (AGI) (Zhan et al., 2025), the creation of general-purpose methods that can efficiently and robustly solve inverse problems constitutes a foundational requirement.

Extensive investigation has been dedicated to addressing the intractable nature of inverse problems. Within the domain of traditional Data Assimilation (DA), a primary strategy to overcome ill-posedness has been the evolution from single-point estimation (e.g., gradient-based and simulation-optimization methods) toward Bayesian inference and ensemble-based approaches. This shift has produced numerous algorithms (Han et al., 2020; Jeong & Park, 2019; Oliver & Chen, 2010; Sayyafzadeh et al., 2012; Yeh et al., 2007; Zhou et al., 2014). Concurrently, addressing the significant challenge of high nonlinearity within inverse modeling has been a major theme, leading to the development of specific notable DA methods like Iterative Local Updating Ensemble Smoother (ILUES) and Ensemble Smoother with Multiple Data Assimilation (ESMDA) (Emerick & Reynolds, 2013; Zhang et al., 2018). However, these conventional algorithms, whether gradient-based, non-gradient-based, simulation-optimization, or ensemble-based, are all constrained by significant computational inefficiency. This limitation stems either from the high cost associated with gradient computations or from the heavy reliance on numerous iterative or ensemble-based calls to the forward model (Wang et al., 2024).

To address the computational bottleneck imposed, a primary strategy involves employing deep learning surrogate models to replace computationally expensive physical simulators (Mo et al., 2019). This composite inversion framework significantly enhances computational efficiency. Furthermore, to mitigate the curse of dimensionality, inversion workflows frequently incorporate an explicit parameterization module. These modules, whether based on traditional statistics (e.g., Karhunen-Loève Expansion) or deep generative models (e.g., Variational Autoencoders (VAEs), Generative Adversarial Networks (GANs), or denoising diffusion probabilistic model (DDPM)), are designed to compress high-dimensional parameter fields into a low-dimensional latent space for modeling, thereby substantially improving efficiency (Canchumuni et al., 2021; Di Federico & Durlofsky, 2025; Zheng et al., 2023). However, the applicability of these methods faces notable limitations across geological settings. Specifically, parameterization strategies effective for Gaussian $K$ fields are often distinct from those with complex non-Gaussian characteristics, such as fluvial channels, turbidites, or preferential flow paths. To accurately characterize the complex spatial structures and geometric morphologies of non-Gaussian geological bodies, the community has specifically developed numerous deep learning-based parameterization methods for integration as discrete modules within inversion frameworks (Han et al., 2022; Mo et al., 2020; Zhan et al., 2022; X. Zhang, 2024; Zhou et al., 2022).

This reliance on task-specific models diverges from the AGI paradigm of integration and generalization. A more streamlined alternative is direct inversion, which involves constructing a single, end-to-end deep learning framework capable of directly learning the mapping from observations $y_{obs}$ to model parameter $m$, thereby circumventing the fragmented development of disparate inversion components (Dai et al., 2025). Several studies have attempted to achieve this goal using GANs or physics-informed neural networks (PINNs) (Fu et al., 2023; Guo et al., 2023; Sun, 2018). Recently, Wang et al. (2024) proposed the physics-informed convolutional decoder (PICD), a significant advancement in direct inversion. PICD leverages a convolutional decoder to approximate hydraulic head fields and employs Karhunen–Loève expansion (KLE) to parameterize hydraulic conductivity fields. The framework simultaneously optimizes decoder parameters and KLE coefficients to minimize observational misfit and physical residuals, with the converged KLE vector representing the estimated hydraulic conductivity field. However, this design yields only a single optimal estimate and cannot circumvent the limitations imposed by linear assumption-based parameterization methods, rendering it applicable primarily to Gaussian scenarios. The underdetermined nature of high-dimensional inverse problems implies that the mapping from sparse $y_{obs}$ to $m$ is non-unique. Consequently, exploring multiple plausible mappings is essential for uncertainty quantification in inverse solutions. In deep learning contexts, uncertainty can be assessed through ensemble methods involving repeated network training with random initialization of weights and biases, which has demonstrated empirical effectiveness (Fort & Jastrzebski, 2019). However, a more direct approach is to leverage the inherent probabilistic properties of deep generative models, particularly diffusion models, which have achieved rapid breakthroughs in both theory and methodology in recent years (Yang et al., 2023).

In this work, we propose a direct probabilistic inversion framework termed Diffusion-Inversion-Net (DIN). The main contributions of this work can be summarized as follows. First, DIN employs DDPM as a powerful geological prior learner. During the training phase, it flexibly incorporates multi-source observational data, including sparse hydraulic head, solute concentration, and hard data $K$, through conditional injection mechanisms. During sampling, these conditioning inputs guide the generative inversion process. Second, DIN leverages stochastic sampling and probabilistic modeling mechanisms to generate multiple constraint-satisfying realizations from identical observational conditions, enabling effective posterior uncertainty quantification without auxiliary design. Third, by exploiting the stable training dynamics and latent space representation capabilities of the diffusion model, DIN directly models posterior distributions regardless of their multimodality or complexity, achieving effective generalization across diverse scenarios (Gaussian or non-Gaussian) for any pixel-based data. Finally, through one-time offline training, DIN amortizes computational costs to the training phase. Once trained, the model rapidly generates posterior samples via the reverse diffusion process, completely bypassing iterative forward simulations and conventional data assimilation procedures. This approach provides a novel and efficient pathway for robust groundwater inversion.

## 2 Methodology

### 2.1 Problem Statement

This study primarily focuses on groundwater flow and solute contaminant transport processes, whose dynamics are governed by the coupling of the steady-state Darcy flow equation and the advection–dispersion equation. Details of the governing partial differential equations are provided in the *Supporting Information*. The numerical solutions obtained from simulators provide the spatiotemporal evolution of the hydraulic head ($h$). When coupled with advection-dispersion equations, these solutions yield the corresponding distribution of contaminant concentration ($C$) within the aquifer. The process of computing model responses from known parameters using these simulators is referred to as forward modeling, which can be expressed as:

$$y_{obs} = \mathcal{A}(m) + \epsilon \tag{1}$$

Here, $\mathcal{A}: \mathbb{R}^{N_m} \mapsto \mathbb{R}^{N_y}$ represents the forward operator, where $N_m$ is the dimension of the parameter vector $m$ and $N_y$ is the dimension of the observation vector $y_{obs}$; $m$ refers to the generalized model parameters; $y_{obs}$ denotes the observed data; and $\epsilon \sim \mathcal{N}(0, \Sigma)$ denotes the observation error vector, which is assumed to follow a Gaussian distribution.

However, the unknown nature of the model parameters $m$ introduces substantial uncertainty into the aforementioned forward model. Characterizing hydraulic conductivity fields represents a typical class of inverse problems, which is the central focus of the present study (Zhang et al., 2024). The fundamental challenge lies in estimating unknown model parameters ($m$) by integrating observed monitoring data with site investigation results and expert knowledge. Both hard ($K$, etc.) and soft data ($h$, $C$, etc.) are the most commonly used data types to solve inverse problems and reveal subsurface structures (Cui et al., 2024). In most cases, solutions to these problems may not exist, may not be unique, or may be highly sensitive to data errors, resulting in ill-posedness. Due to the ill-posed nature of the problem, infinitely many feasible solutions exist, and perfect recovery is impossible (Tarantola, 2005).

Building upon this foundation, uncertainty quantification is essential (Kitanidis, 2015). Traditional inversion relies on extensive forward simulations for uncertainty quantification. We focus on end-to-end probabilistic generative inverse solvers that eliminate repeated forward computations while enabling efficient uncertainty quantification.

### 2.2 Denoising diffusion probabilistic model for prior-learning

DDPMs, originally introduced by Sohl-Dickstein et al. (2015), are a class of deep generative models designed to learn and approximate complex data distributions, enabling the generation of realistic samples therefrom. In the context of inverse modeling, the DDPM can be regarded as a powerful prior-learning framework that captures the intrinsic diversity of spatially heterogeneous hydraulic conductivity fields ($m_0$), whether Gaussian or non-Gaussian in nature.

The original DDPM consists of two components: forward diffusion process $q(m_t \mid m_{t-1})$ and reverse denoising process $p_\theta(m_{t-1} \mid m_t)$, as shown in Figure 1 (a). Here we consider DDPM as stochastic Markov chains made up of $T$ steps. The sequence of step size is controlled by a variance schedule $\{\beta_t \in (0,1)\}_{t=1}^T$, typically following a linear monotonically increasing sequence (Ho et al., 2020). Conceptually, the DDPM learns to invert a gradual Gaussian noising process, allowing complex geologic priors to be represented as simple Gaussian distributions in latent space. The $q(m)$ can be characterized using the following equation:

$$q(m_t \mid m_{t-1}) = \mathcal{N}(m_t; \sqrt{1-\beta_t} m_{t-1}, \beta_t I) \tag{2}$$

$$q(m_{1:T} \mid m_0) = \prod_{t=1}^{T} q(m_t \mid m_{t-1}) \tag{3}$$

where $m_t$ denotes the latent variable at diffusion step $T$. Provided that $T$ is sufficiently large and a well-structured schedule of $\beta_t$ is employed, the resulting $m_T$ tends to approximate an isotropic Gaussian distribution. Consequently, if the precise knowledge of the reverse distribution $q(m_{t-1} \mid m_t)$ is possessed, we can sample $m_T \sim \mathcal{N}(0, I)$ and execute the process in reverse to obtain a sample from $q(m_0)$. However, given that $q(m_{t-1} \mid m_t)$ relies on the entirety of the data distribution, it is approximated through the utilization of a neural network in the following manner:

$$p_\theta(m_{t-1} \mid m_t) = \mathcal{N}\big(m_{t-1}; \mu_\theta(m_t, t), \Sigma_\theta(m_t, t)\big) \tag{4}$$

$$p_\theta(m_{0:T}) = p(m) \prod_{t=1}^{T} p_\theta(m_{t-1} \mid m_t) \tag{5}$$

The symbol $\theta$ signifies the parameters that the designed network architecture needs to learn. Using mathematical calculations and reparameterization trick, Bayesian formula and the properties of Markov chains, the posterior mean ($\mu_\theta$) and variance ($\Sigma_\theta$) can be expressed as functions of $m_t$ and $t$.

$$\Sigma_\theta = \frac{1 - \bar{\alpha}_{t-1}}{1 - \bar{\alpha}_t} \beta_t \tag{6}$$

$$\mu_\theta = \frac{\sqrt{\alpha_t}(1 - \bar{\alpha}_{t-1})}{1 - \bar{\alpha}_t} m_t + \frac{\sqrt{\bar{\alpha}_{t-1}} \beta_t}{1 - \bar{\alpha}_t} m_0 \tag{7}$$

Here, $\alpha_t = 1 - \beta_t$, $\bar{\alpha}_t = \prod_{i=1}^{t} \alpha_i$ are defined for simplification purpose. To learn a valid prior distribution and produce genuine data samples $m_0$, the reverse denoising process should undergo training with the goal of maximizing the probability distribution $p_\theta(m_0)$, which is equivalent to minimizing the negative log-likelihood, $-\log p_\theta(m_0)$. Ho et al. (2020) demonstrated that the model parameters $\theta$ can be optimized by minimizing the negative log-likelihood through a variational lower bound. From this formulation, the simplified training objective $L_{simple}^{DDPM}$—which omits the weighting term involving $\beta_t$ and $\Sigma_\theta$ in diffusion models—can be derived as follows:

$$\begin{aligned} L_{simple}^{DDPM} &= \mathbb{E}_{t \sim [1,T], m_0, \epsilon_t}\big[\| \epsilon_t - \epsilon_\theta(\sqrt{\bar{\alpha}_t} m_0 + \sqrt{1 - \bar{\alpha}_t} \epsilon_t, t) \|^2\big] \\ &= \mathbb{E}_{t \sim [1,T], m_0, \epsilon_t}\big[\| \epsilon_t - \epsilon_\theta(m_t, t) \|^2\big] \end{aligned} \tag{8}$$

Where $\epsilon_t \sim \mathcal{N}(0, I)$ is a Gaussian noise at any time-step, and $\epsilon_\theta$ is the noise predictor operating with inputs $\sqrt{\bar{\alpha}_t} x_0 + \sqrt{1 - \bar{\alpha}_t} \epsilon_t$ and $t$, whose objective is to predict the noise added during the forward process.

### 2.3 Diffusion-Inversion-Net for inverse modeling

A robust inversion framework must incorporate two essential components: 1) a realistic prior distribution that characterizes the spatial heterogeneity of the subsurface property, and 2) a data-consistency mechanism that ensures the generated fields are compatible with the available observations. The newly proposed end-to-end probabilistic inversion solver, DIN, integrates these two components within a unified generative–inversion paradigm. Specifically, the DDPM provides a learned prior over the model-parameter space, while conditional information is introduced during the denoising process to steer the sampling trajectory toward solutions that are both consistent with the observations and geologically plausible.

DIN adopts a classifier-free guidance (CFG) approach to achieve stable conditioning during spatial prior learning. Specifically, the conditional inputs are randomly dropped with a fixed probability $p$, enabling the network to learn both the unconditional noise predictor $\epsilon_\theta(m_t, t)$ and the conditional one $\epsilon_\theta(m_t, t, y_{obs})$ within a single training run. During inversion process, conditional guidance is applied by linearly combining the two noise predictions:

$$\tilde{\epsilon}_\theta = \epsilon_\theta(m_t, t) + \lambda\big(\epsilon_\theta(m_t, t, y_{obs}) - \epsilon_\theta(m_t, t)\big) \tag{9}$$

where $\lambda$ is a user-defined guidance scale that balances geological realism and observation consistency. This formulation allows the conditional signal to steer the denoising trajectory without retraining or altering the network architecture.

While CFG provides stable conditional guidance, additional constraints are needed to explicitly enforce data consistency at the observation level. To better exploit the physical observations in field inversion, the loss function of DIN is augmented with an observation alignment term. After predicting the noise $\epsilon_\theta(m_t, t, y_{obs})$ at each diffusion step, the corresponding denoised realization $m_o^{(t)}$ is estimated using the standard DDPM reparameterization. The simulated observation $y_{obs}^{(t)} = \mathcal{A}(x_o^{(t)})$ is then compared with the measured observation $y_{obs}$. The alignment loss is formulated as:

$$L_{align}^{DIN} = \mathbb{E}_t\left[\left\|F\big(m_o^{(t)}\big) - y_{obs}\right\|^2\right] \tag{10}$$

and added to the conventional denoising objective:

$$L_{total}^{DIN} = L_{simple}^{DIN} + \eta L_{align}^{DIN} \tag{11}$$

$$= \mathbb{E}_{t\sim[1,T],m_0,\epsilon_t}[\|\epsilon_t - \epsilon_\theta(m_t, t, y_{obs})\|^2] + \eta\mathbb{E}_t\left[\left\|F(m_o^{(t)}) - y_{obs}\right\|^2\right]$$

where $\eta$ controls the trade-off between distributional fidelity and observation consistency. This dual-loss formulation ensures that the learned prior remains statistically faithful to the training datasets, while simultaneously enforcing agreement with measured data.

Once trained, DIN performs inversion by sampling an initial Gaussian noise $m_T \sim \mathcal{N}(0, I)$ and progressively denoising through the conditional guidance described above. At each step, the guided noise $\tilde{\epsilon}_\theta$ is used to estimate $m_{t-1}$ following the DDPM posterior update. The resulting samples are thus both geologically realistic and dynamically consistent with field observations.

A significant advantage of the DIN is its inherent capability for uncertainty quantification. By leveraging the stochastic nature of the reverse diffusion process, the model performs implicit Bayesian inference to approximate the posterior distribution $\mathcal{P}(m_0 \mid y_{obs})$. Repeating the conditional sampling process $N_e$ times yields an ensemble of plausible realizations, $\mathcal{M} = \left\{m_0^{(1)}, m_0^{(2)}, \ldots, m_0^{(N_e)}\right\}$. This ensemble provides a robust basis for statistically analyzing the uncertainty associated with the inversion results, moving beyond a single deterministic estimate.

The above DIN framework can be implemented with any backbone network capable of noise prediction. In this study, a modified U-Net architecture was employed, and the architectural details are provided in the *Supporting Information*.

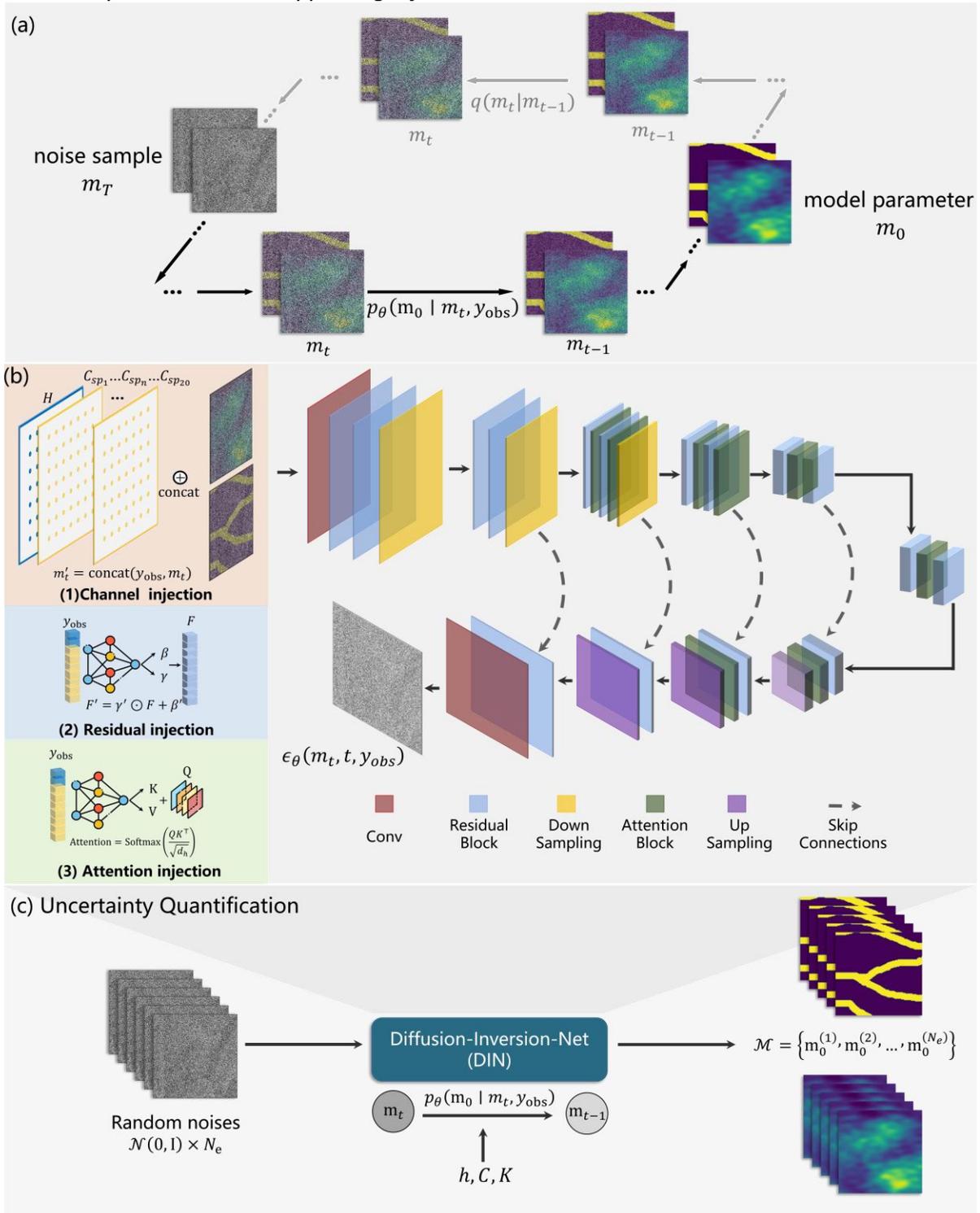

Fig. 1 (a) Schematic diagram of the DIN framework. (b) The U-Net architecture employed by DIN, illustrating the three conditional injection strategies. (c) Schematic illustration of the uncertainty quantification process using DIN.

## 3 Dataset

### 3.1 Gaussian Case

The model DIN operates within a data-driven framework that necessitates a substantial amount of paired data, comprising model parameters $m$ and groundwater simulation model responses (i.e., observation data $y_{obs}$). For the Gaussian case, all hydraulic conductivity fields were defined on a 128 × 128 grid. A large ensemble of hydraulic conductivity realizations, representing the model parameters of interest $m$ (with a dimension of 16,384), was randomly generated using the KLE. These Gaussian hydraulic conductivity fields were then used as inputs to MODFLOW and MT3DMS to simulate groundwater flow and solute transport processes. Hydraulic head and solute concentration data were collected at multiple time steps (stress periods) and observation wells, and combined with hard data of hydraulic conductivity at the same locations to form observation vectors $y_{obs}^{sys}$, with a dimension of 1,078. Although this random generation strategy is relatively straightforward, it yielded a diverse training dataset encompassing a wide range of $m \sim y_{obs}^{sys}$ relationships.

### 3.2 Non-Gaussian Case

The conceptualization of the field and the construction of the dataset for this case follow the same approach as in the Gaussian case, with the only difference being in the generation of the hydraulic conductivity field. In this case, the hydraulic conductivity fields were randomly extracted from the training image (TI) by cropping the TI using a 128 × 128 kernel (Laloy et al., 2018). To ensure that the global statistical characteristics of the TI is adequately represented in our dataset, the commonly used Latin Hypercube Sampling (LHS) method was employed (Zhang & Pinder, 2003). Following the aforementioned process, two datasets were constructed, each containing 5,000 unique pairs of hydraulic conductivity $m$ and corresponding observed data $y_{obs}^{sys}$, one based on a Gaussian distribution and the other on a non-Gaussian distribution. Additionally, a separate set of 200 independently generated testing pairs was created to assess the performance of the network.

The complete conceptualization, detailed configuration, and algorithmic principles of the KLE for this numerical case, along with the dataset preparation procedures, are provided in the *Supporting Information*.

## 4 Results

### 4.1 Effect of Conditional Injection Strategies

In conditional DDPMs, the mechanism for incorporating conditional information critically influences generative accuracy and stability. Existing studies have explored multiple injection strategies—channel concatenation (Kazemi & Esmaeili, 2025), residual modulation (Perez et al., 2017), and attention-based fusion (Park et al., 2025; Rombach et al., 2022)—each offering complementary benefits.

In the present study, the DIN framework employs a DDPM backbone to learn the geological prior while investigating how the hierarchical integration of conditioning pathways influences the inversion accuracy. To systematically examine this effect, three progressively enriched conditioning schemes were designed: (1) Channel injection (2) Channel + Residual injection, and (3) Channel + Residual + Attention injection. The three conditioning injection strategies are provided in the *Supporting Information*.

To ensure methodological consistency while maintaining computational efficiency, each configuration was trained three times on 500 Gaussian samples and evaluated through 200 independent realizations, with RMSE and SSIM (metrics illustrated in *Supporting Information*). The comparative results (Figure 2 (a)), demonstrate a clear, monotonic improvement across the three conditioning schemes. Specifically, RMSE values exhibit a consistent decline, indicating that the inversion estimates become increasingly accurate overall. Although the variance increases and the interquartile range expands from (1) channel to (3) channel + residual + attention injection, the overall SSIM scores correspondingly improve. This suggests that more sophisticated conditioning improves performance, though limited training data likely causes the increased uncertainty. To determine the optimal training dataset size for DIN, we conducted a series of experiments using a Gaussian case with sample sizes of $N \in \{500, 1000, 2000, 3000, 5000\}$. From this analysis, 5,000 samples were identified as the optimal size. These results are presented in Figure 2(b), and a detailed discussion is provided in the *Supporting Information*.

Furthermore, any observational information that contributes to the inversion of subsurface structures (e.g., geophysical data) can be incorporated into the U-net network through any of the aforementioned conditional injection strategies.

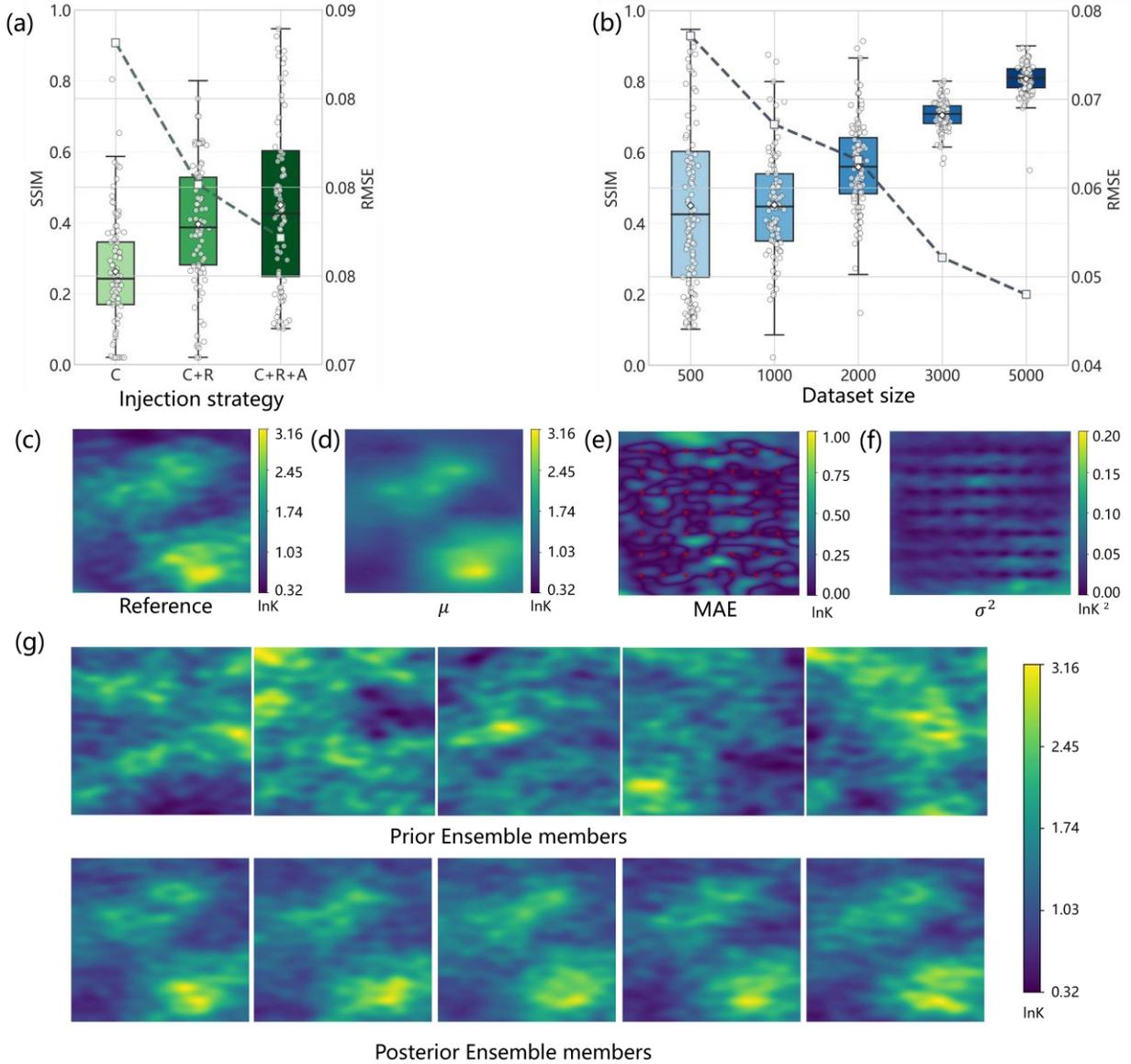

Fig. 2 Performance evaluation, based on (a) SSIM and (b) RMSE metrics, for different conditional injection strategies and varying training dataset sizes. (c)-(g) illustrate the generative inversion results from the DIN framework for the Gaussian case.

### 4.2 Inversion Results for the Gaussian Case

The DIN model was trained using a dataset comprising 5,000 paired samples, each consisting of model parameters ($m$) and corresponding model responses ($y_{obs}^{sys}$) from 49 monitoring wells, which include both hard data ($K$) and soft data ($C, H$). In the inversion stage, the trained DIN model is conditioned on real observations $y_{obs}^{real}$ from a single reference field, as illustrated in Figure 2 (c).

As described in Section 2.4, the probabilistic generative nature of the DDPM enables a natural representation of posterior uncertainty quantification. Specifically, the well-trained DIN model was executed independently for $N_e = 100$ sampling runs, generating a set of posterior ensemble samples consistent with the observed field data $y_{obs}^{real}$. Based on this ensemble, the posterior mean $\mu$ and posterior variance $\sigma^2$ were computed and are presented in Figures 2 (d) and (f), respectively. The mean absolute error (MAE) between the $\mu$ and the reference field is shown in Figure 2 (e). Furthermore, to demonstrate the shift from prior to posterior and the diversity of the posterior realizations, 5 random samples from the prior and posterior ensembles are compared in Figure 2(g).

From the posterior mean $\mu$, it is evident that the DIN model can accurately estimate the high-dimensional Gaussian hydraulic conductivity field using a low-dimensional observations. The spatial distribution characteristics of both high- and low-permeability zones in the reference field are well captured in the posterior mean. Meanwhile, the MAE results reveal the presence of several filament-like low-error bands in the error distribution, indicating that the posterior ensemble exhibits relatively high estimation confidence in these regions. Notably, the fine-scale geological structures corresponding to these filament-like low-error bands are precisely recovered, which fully demonstrates the DIN model's remarkable capability in generating high-quality samples and capturing high-frequency detailed features. This characteristic can be attributed to its progressive denoising mechanism, which enables the model to continuously refine image details during the step-by-step denoising process.

Furthermore, we present the dynamic evolution of posterior samples throughout the 1000-step denoising process, as illustrated in the video provided in the *Supporting Information*. The video demonstrates that during the denoising process, the DIN model first reconstructs the overall low-frequency structures (such as the distribution patterns of high- and low-permeability regions), followed by the gradual recovery of high-frequency details (such as local edge morphologies and texture). This hierarchical generation pattern is consistent with findings reported in existing theoretical studies on diffusion models (Falck et al., 2025).

It can be observed that these filament-like low-error bands spatially connect the locations of the 49 monitoring wells (indicated by red dots in Figure 2 (e)), while regions away from the monitoring wells exhibit a certain degree of error in the inversion results due to sparse observational information. We attribute this phenomenon to the unique channel concatenation conditional injection strategy of the DIN model. This mechanism implements conditional constraints by embedding observational data into the channels in the form of spatial coordinates (see Supporting Information for details). Specifically, in regions with higher MAE values, the constraints imposed by observational data are relatively weak, and the model relies more heavily on the learned prior distribution of Gaussian hydraulic conductivity fields from KLE.

As shown in Figure 2 (f), the posterior variance $\sigma^2$ indicates that the value is significantly lower at monitoring well locations, demonstrating that the hard data $K$ at these locations has been effectively learned by the DIN model, which imposes strong local constraints and reduces posterior ensemble estimates uncertainty at monitoring well locations. Concurrently, the results presented in Figures 2 (f) and (g) collectively demonstrate that the 100 posterior samples maintain necessary diversity while keeping uncertainty within a reasonable range, thus enabling reliable quantification of uncertainty in the Gaussian case.

Moreover, quantitative metrics indicate that the RMSE and SSIM values calculated from the $N_e = 100$ posterior realizations are 0.042 and 0.85, respectively.

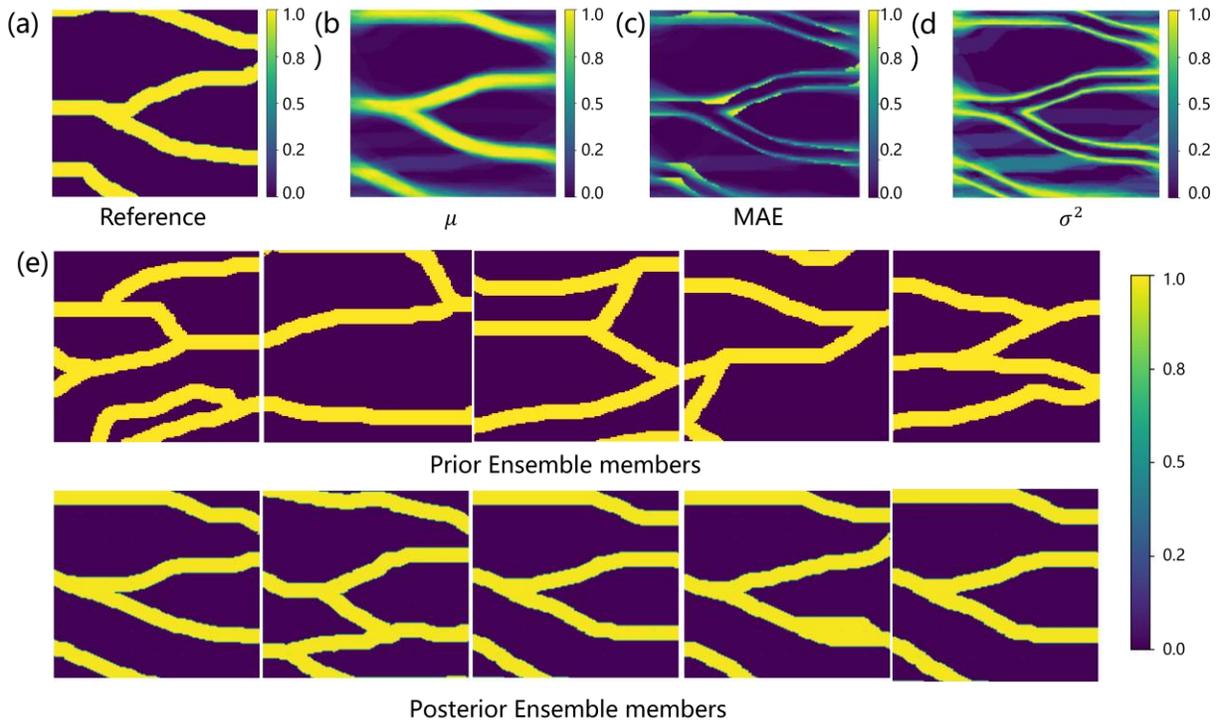

Fig. 3 Generative inversion results from the DIN framework for the non-Gaussian case.

### 4.3 Inversion Results for the Non-Gaussian Case

In practical applications, the characterization of non-Gaussian aquifer structures has also received considerable attention (Zhang et al., 2024), and compared to the Gaussian case, the corresponding inverse problem is more challenging.

Similar to the Gaussian case, we employed the DIN model pre-trained on the non-Gaussian dataset to perform $N_e$ generative inversion for a single reference field (Figure 3 (a)), using its corresponding real observations $y_{obs}^{real}$. To systematically demonstrate the DIN model's inversion performance, we present the mean $\mu$ and variance $\sigma^2$ of the posterior samples, the MAE results, as well as 5 randomly selected prior and posterior ensemble members, repectively.

From the posterior mean $\mu$, it can be observed that the DIN model demonstrates exceptional capability in the generative inversion of non-Gaussian hydraulic conductivity fields. Although errors exist in certain regions (as indicated by the MAE), the channel features in most posterior samples exhibit strong consistency with that in the reference field. The samples provided in the *Supporting Information* further corroborates the generation pattern consistent with the Gaussian case, wherein low-frequency structures are reconstructed first, followed by progressive recovery of high-frequency details.

The posterior variance $\sigma^2$ reveals that uncertainty is relatively higher at the boundaries between high- and low-permeability zones. This phenomenon is further evidenced by the results of selected posterior ensemble members: while maintaining consistency in the overall low-frequency structures, different samples exhibit certain variations in high-frequency details, with artifacts appearing in boundary regions primarily attributable to incomplete recovery of high-frequency features during the denoising process. Nevertheless, the uncertainty arising from this incomplete high-frequency recovery remains within an acceptable range, indicating that the DIN model is still capable of achieving robust and reliable generative inversion and uncertainty quantification under non-Gaussian cases. Quantitative evaluation metrics were also computed for the inversion results of the sample shown in Figure 3 (e). The RMSE and SSIM values derived from the posterior samples were 0.077 and 0.90, respectively.

Unlike deterministic end-to-end inversion approaches, DIN's denoising diffusion-based inversion generates a substantially richer ensemble of solutions, preventing mode collapse and overconfident predictions. Our DIN model not only produces a single reasonable hydraulic conductivity field estimate but also captures the complete posterior distribution over all feasible fields, thereby enhancing the robustness of decision-making processes in groundwater modeling.

## 5 Discussion and conclusion

We introduce DIN, a direct probabilistic inversion solver based on conditional generative modeling, specifically designed for inferring hydraulic conductivity fields in groundwater flow and solute transport systems. Across both Gaussian and non-Gaussian cases, DIN demonstrates robust inversion performance, offering a compelling alternative to conventional iterative or deterministic inversion frameworks.

DIN's performance stems from three core strengths: (a) its efficacy in resolving the end-to-end inversion from low-dimensional, sparse observational data to high-dimensional model parameters; (b) its capacity to overcome challenges posed by strong subsurface heterogeneity, achieving effective generalization across diverse geostatistical distributions ; and (c) its integration of flexible conditional injection mechanisms, which, combined with its inherent probabilistic generative nature, naturally enables effective uncertainty quantification .

We also note concurrent research in surface water data assimilation, which utilizes a Latent Diffusion Model embedded within an ensemble assimilation framework to perform assimilation in latent space (Foroumandi & Moradkhani, 2025). Notably, the approach employs score-based diffusion modeling, whereas our DIN adopts a noise-prediction formulation. Although mathematically equivalent under appropriate reparameterization (Vincent, P., 2011; Song & Ermon, 2020), the noise-prediction objective may yield more stable and efficient training in practice. This complementary line of research further reinforces the broader conclusion that generative diffusion models provide a powerful, flexible, and scalable foundation for inverse problems in Earth system sciences.

It should be noted that the current implementation trains separate models for distinct prior distributions (Gaussian and non-Gaussian), limiting generalization to unseen parameter distributions. Nevertheless, we are confident that this framework can be extended to jointly model heterogeneous conductivity fields with mixed statistical characteristics within a unified architecture. Moreover, fine-tuning a pre-trained DIN on site-specific data offers a practical and efficient adaptation strategy. A particularly promising direction is the development of a foundation model pre-trained on large-scale (Cao et al., 2025), diverse geological datasets, which would enable rapid deployment through transfer learning for real-world applications in hydrogeology, environmental engineering, and beyond.

**Acknowledgments**

This work was funded by National Natural Science Foundation of China (42572309). Many thanks to the anonymous reviewers whose comments have helped to improve this manuscript.

**Conflict of Interest**

The authors declare no conflicts of interest relevant to this study.

**Open Research**

The source code for DIN is available at https://github.com/RaphaelYangWJ/Diffusion-Inversion-Net and is permanently archived at https://doi.org/10.5281/zenodo.17642304.